# Dimensionality crossover to 2D vestigial nematicity from 3D zigzag antiferromagnetism in an XY-type honeycomb van der Waals magnet

Zeliang Sun[1,+], Gaihua Ye[2,+], Mengqi Huang[3,+], Chengkang Zhou[4+], Nan Huang[5,6], Qiuyang Li[1], Zhipeng Ye[2], Cynthia Nnokwe[2], Hui Deng[1], David Mandrus[5,6], Zi Yang Meng[4], Kai Sun[1], Chunhui Du[3], Rui He[2,#], Liuyan Zhao[1,#]

[1] Department of Physics, University of Michigan, Ann Arbor, MI 48109, USA
[2] Department of Electrical and Computer Engineering, Texas Tech University, Lubbock, TX 79409, USA
[3] School of Physics, Georgia Institute of Technology, Atlanta, GA 30332, USA
[4] Department of Physics and HKU-UCAS Joint Institute of Theoretical and Computational Physics, The University of Hong Kong, Pokfulam Road, Hong Kong SAR, China
[5] Department of Materials Science and Engineering, The University of Tennessee, Knoxville, TN 37996, USA
[6] Materials Science and Technology Division, Oak Ridge National Laboratory, Oak Ridge, TN 37831, USA

[+] These authors contributed equally.
[#] Corresponding authors: lyzhao@umich.edu (L.Z.); rui.he@ttu.edu (R.H.)

**Fluctuations and disorder effects are substantially enhanced in reduced dimensionalities[1,2]. While they are mostly considered as the foe for long-range orders, fluctuations and disorders can also stimulate the emergence of novel phases of matter, for example, vestigial orders[3,4]. Taking two-dimensional (2D) magnetism as a platform, existing efforts have been focused on maintaining 2D long-range magnetic orders by suppressing the effect of fluctuations[5-9], whereas the other side, exploiting fluctuations for realizing new 2D magnetic phases, remains as an uncharted territory. Here, using a combination of nitrogen-vacancy (NV) spin relaxometry, optical spectroscopy, and Monte Carlo simulations, we report, in an XY-type honeycomb magnet $NiPS_3$, the phase transition from the zigzag antiferromagnetic (AFM) order in three-dimensional (3D) bulk to a new $Z_3$ vestigial Potts-nematicity in 2D few layers. Spin fluctuations are shown to significantly enhance over the GHz–THz range as the layer number of $NiPS_3$ reduces, using the NV spin relaxometry and the optical Raman quasi-elastic scattering. As a result, the Raman signatures of the zigzag AFM for bulk $NiPS_3$, a zone-folded phonon at ~30cm$^{-1}$ from the broken translational symmetry ($P_{BTS}$) and a degeneracy lift of two phonons at ~180cm$^{-1}$ for the broken 3-fold rotational symmetry ($P_{BRS}$), evolve into the disappearance of $P_{BTS}$ and the survival of $P_{BRS}$ in few-layer $NiPS_3$, with a critical thickness of ~10nm. The optical linear dichroism microscopy images all three nematic domain states in a single few-layer $NiPS_3$ flake. The large-scale Monte Carlo simulations for bilayer $NiPS_3$ model confirms the absence of long-range zigzag AFM order but the formation of the $Z_3$ vestigial Potts-nematic phase, which corroborates with the experimental finding of 3-fold rotational symmetry breaking, but translational symmetry restoring in <10nm $NiPS_3$. Our results demonstrate the positivity of strong quantum fluctuations in creating new phases of matter after destroying more conventional ones, and thus, offer an unprecedented pathway for developing novel 2D phases.**

A vestigial order describes the partial melting of the primary order ($\eta$) of a spontaneous symmetry breaking phase[3,4]. In contrast to conventional short-range orders that do not break symmetries of their hosting crystalline lattices, a vestigial order corresponds to a composite order ($\eta^\dagger \tau \eta$, a quadratic form of $\eta$) and breaks a subset of symmetries broken by the primary order[3,4]. Fluctuations and disorders are the typical causes for destroying primary orders, and potentially, introduce vestigial orders. Illustrated in Fig. 1a, the increasing temperature provides thermal fluctuations, and the reduced dimensionality can independently contribute to enhanced fluctuations. In between the long-range primary ordered region I ($\langle\eta\rangle \neq 0$ and $\langle\eta^\dagger \tau \eta\rangle \neq 0$) and the completely disordered region III ($\langle\eta\rangle = 0$ and $\langle\eta^\dagger \tau \eta\rangle = 0$), there could exist the region II for the vestigial ordered case[3,4] ($\langle\eta\rangle = 0$ but $\langle\eta^\dagger \tau \eta\rangle \neq 0$). The impact of introducing vestigial order has been profound, as it provides the intertwined nature of proximate phases within the rich phase diagrams for many quantum material systems including Cu-based high-$T_c$ superconductors[3,10-14], Fe-based superconductors[15-18], charge density wave systems[19] and others[20]. Theoretical examples of primary versus vestigial orders are ample in many systems: charge or spin density wave v.s. nematicity[4,21,22], pair density wave v.s. charge-4$e$ superconductivity or nematicity or charge density wave[23-26], superconductivity v.s. charge-4$e$ superconductivity or nematicity[27-29], and even the anyon condensation from $Z_2 \times Z_2$ topological order quantum spin liquid v.s. $Z_2$ topological order quantum spin liquid[30], *etc*.

Yet, experimental realizations of vestigial order have been primarily focused on the nematicity that breaks the rotational symmetry but preserves the translational symmetries of underlying crystal lattices[4,22]. This symmetry requirement allows the tetragonal and hexagonal systems to support the emergence of nematicity. Indeed, in tetragonal systems such as cuprates[13,14,31,32] and pnictides[16,22,33,34], $Z_2$ Ising-nematicity is observed to develop from the charge and spin fluctuations, which breaks the tetragonal 4-fold into the orthorhombic 2-fold rotational symmetry and retains the translation symmetries. In contrast, in hexagonal systems, $Z_3$ Potts-nematicity is anticipated, which breaks the hexagonal 6-fold (or 3-fold) rotational symmetry and preserves the translational symmetries[35-37]. Very recently, in systems such as Fe$_{1/3}$NbS$_2$[38] and FePS$_3$[39], characters of Potts-nematicity have been seen in their long-range ordered magnetic phases, i.e., region I in Fig. 1a. However, the intrinsic Potts-nematic order without the primary order, i.e., region II in Fig. 1a, has been much less explored.

Reducing dimensionality is one promising, but less explored route to enhance fluctuations, and in fact, pioneering theoretical studies on nematicity often used two-dimensional (2D) models[4]. In this regard, we choose a 2D honeycomb magnetic system with the XY-type spin anisotropy, NiPS$_3$[40-44], to harvest stronger spin fluctuations for realizing $Z_3$ Potts-nematicity. This study makes a distinct effort from the mainstream research on 2D magnetism that works on maintaining long-range magnetic orders against fluctuations[5-9]. The monolayer NiPS$_3$ has the trigonal crystallographic point group $D_{3d}$ with an out-of-plane 3-fold rotational axis, and the bulk and few-layer NiPS$_3$ obey the monoclinic structural point group $C_{2h}$ due to the lateral shift between adjacent layers along the $a$ axis. The bulk NiPS$_3$ undergoes an antiferromagnetic (AFM) transition at $T_{AFM,3D}$ = 155K, below which the Ni$^{2+}$ spins align along the $a$ axis ferromagnetically (FM) within the zigzag chains and are coupled AFM between neighboring chains[45-49] (Fig. 1b). Such a long-range zigzag AFM order in NiPS$_3$ spontaneously breaks the 3-fold rotational symmetry of individual layers, and at the same time, breaks the translational symmetry with a wavevector $\mathbf{Q} = \mathbf{k_M}$ where $\mathbf{k_M}$ is the momentum at the M point in the Brillouin zone (Fig. 1b inset). Three degenerate zigzag AFM states are expected, with three $\mathbf{Q}$ rotated from each other by 120°, i.e., $\mathbf{Q}_j$ (j = 1, 2, 3). Such a broken rotational symmetry (BRS) and broken translational symmetry (BTS) AFM phase in bulk NiPS$_3$ has a non-zero primary order parameter $\eta_j(\mathbf{r}) = \big(\mathbf{S}(\mathbf{r}) - \mathbf{S}(\mathbf{r} + \mathbf{e}_j)\big) e^{i\mathbf{Q}_j \cdot \mathbf{r}}$ and also supports a finite secondary order parameter $\eta^\dagger \tau \eta = \left(\frac{\sqrt{3}}{2}(\eta_3^2 - \eta_2^2), \frac{1}{2}(2\eta_1^2 - \eta_2^2 - \eta_3^2)\right)$, belonging to region I in Fig. 1a. Due to the weak anisotropy in the XY-model, strong spin fluctuations in the 2D, and FM-AFM exchange competitions, few-layer NiPS$_3$ is expected to realize the scenario in region II where the nematic phase has BRS but no BTS (Fig. 1c), before going into the completely spin disordered, paramagnetic phase of region III (Fig. 1d).

We start by verifying the enhancement of spin fluctuations in thinner NiPS$_3$ samples, using both the nitrogen-vacancy (NV) spin relaxometry and the quasi-elastic scattering (QES) in Raman spectroscopy that cover distinct frequency regimes from GHz to THz. NV spin relaxometry utilizes the excellent magnetic field sensitivity of NV spin centers to probe fluctuating magnetic fields at the NV electron spin resonance frequencies (~ GHz) generated by proximal magnetic materials[50-54]. The measured NV spin relaxation rate is proportional to the spectral density of the magnetic field noise, which scales with the spin fluctuation strength[52] (see Supplementary Information Note 1). Here, we spatially visualize the spin fluctuations in exfoliated NiPS$_3$ flakes using the wide-field imaging mode of NV spin relaxometry[55]. A representative NV spin relaxation rate map is shown in Fig. 2a inset, for a 4-layer (4L) and a 10nm-thick NiPS$_3$ flake at 170K. Figure 2a shows the temperature dependence of the area-averaged NV spin relaxation rate for the 4L and 10nm NiPS$_3$ flakes, displaying a decreasing trend for spin fluctuations at lower temperatures for both samples. To account for the thickness difference of spin fluctuation, we extract the magnetic susceptibility $\chi$ of 4L and 10nm NiPS$_3$ samples, from the NV spin relaxation rate (see Supplementary Information Note 1). Figure 2b plots $\chi$, which is proportional to the spin fluctuations[56], as a function of temperature for the 4L and 10nm NiPS$_3$ samples. It is evident that $\chi$ for 4L NiPS$_3$ is significantly enhanced in comparison with that for 10nm NiPS$_3$ below ~ 150K, whereas it is only slightly larger in 4L than in 10nm sample above ~ 150K.

Complementary, Raman QES in NiPS$_3$ arises from spin fluctuations[57-61] and manifests as a slow delay profile in the Raman spectra for up to ~ 40cm$^{-1}$ (in the order of THz), with two exemplary Raman spectra of QES for a 4L and a 13.1nm NiPS$_3$ flake taken at 170K shown in Fig 2c. Note that both spectra are normalized to their corresponding thicknesses for a fair comparison of spin fluctuations between the two samples. It is evident that the spin fluctuations are stronger in 4L than in 13.1nm NiPS$_3$ over a spectral range up to ~40cm$^{-1}$, despite the overall decay at higher frequencies for both samples. Moreover, we can integrate the QES spectral weight (SW$_{QES}$) over 8 – 40cm$^{-1}$, with that from the phonon breathing mode excluded, to quantify the spin fluctuations, and track its temperature dependence. Figure 2d shows the temperature dependence of SW$_{QES}$ for bilayer (2L), trilayer (3L), 4L, 13.1nm, and bulk NiPS$_3$, with each trace normalized to its corresponding SW$_{QES}$ at 200K. There is a similarity amongst the SW$_{QES}$ ($T$) plots that QES decreases as temperature reduces with a small kink at a critical temperature around 155K for bulk ($T_{AFM,3D}$) and around 120K for 2L ($T_{N,2D}$). But what is more striking and more important is the contrasting behaviors of QES at low temperatures among NiPS$_3$ of distinct thicknesses: bulk NiPS$_3$ shows a fully suppressed QES at temperatures below a characteristic temperature $T_{sQES}$ ~90K that is lower than $T_{AFM,3D}$, 13.1nm NiPS$_3$ reveals a lower $T_{sQES}$ of 50K, whereas 2L, 3L, and 4L NiPS$_3$ all feature finite QES down to our lowest temperature 10K, i.e., $T_{sQES}$, if present, is lower than 10K. Such a stronger contrast at lower temperatures of spin fluctuations between bulk and few layers is suggestive of different magnetic phases between 3D and 2D NiPS$_3$.

Up to now, it is convincingly shown that as NiPS$_3$ thickness reduces from 3D bulk to 2D flake, spin fluctuations are clearly enhanced, fulfilling the key requirement for realizing vestigial nematicity in 2D NiPS$_3$. A natural next step is to examine the magnetic phases upon thinning 3D bulk NiPS$_3$ down to 2D films, with a proper experimental tool to simultaneously detect BRS and BTS, as these two together are required to distinguish zigzag AFM, Potts-nematicity, and paramagnetism illustrated in Figs. 1b-1d. Here, we select Raman spectroscopy with which BRS can be detected through splitting of degenerate phonons and BTS can be probed by zone-folding phonon modes. Figure 3a shows Raman spectra for bulk NiPS$_3$ taken in linearly parallel and crossed channels at 10K with a 633nm excitation laser, over selected frequency ranges of interest (see full-range Raman spectra in Supplementary Information Note 2). Figures 3b and 3c show the temperature dependent Raman spectra in the linearly parallel channel of Fig. 3a. First, the two Raman modes around 180cm$^{-1}$ are degenerate above $T_{AFM,3D}$ and split below $T_{AFM,3D}$, whose frequency separation was assigned as the signature for zigzag AFM previously[43]. Despite the monoclinic stacking between honeycomb layers, the phonon modes obey the selection rules of D$_{3d}$ above $T_{AFM,3D}$, and the two phonon modes at ~180cm$^{-1}$ belong to the E$_g$(D$_{3d}$) symmetry. It is BRS, rather than BTS, from the zigzag AFM order that lifts the E$_g$(D$_{3d}$) doublet into one A$_g$(C$_{2h}$) and one B$_g$(C$_{2h}$) mode (thus named as P$_{BRS}$). As a result, the frequency separation here shall be proportional to the composite nematic order parameter $\eta^\dagger \tau \eta$, instead of $\eta$. Second, the Raman mode at ~30cm$^{-1}$, emerging below $T_{AFM,3D}$, has not been reported previously[43]. We assign this mode to be a phonon mode folded from the Brillouin zone boundary M point to the center Γ point due to the zigzag AFM order, after the following examinations. We first observed the in-plane and out-of-plane magnetic field independence for this mode and rule out the possibility of magnon[46,48] (see Supplementary Information Note 3). We then compared to the calculated phonon dispersion spectra of bulk NiPS$_3$ and found no intralayer or interlayer Γ point phonons around 30cm$^{-1}$ [62,63]. Considering that the zigzag AFM order has a wavevector **Q** = **k**$_M$, we checked the phonon spectra at the M point, and indeed, found a computed M point phonon mode at ~30cm$^{-1}$ [62]. Furthermore, we examined the polarization dependence of this Raman mode and observed its clear anisotropy, confirming it is a single-Q zone-folding process that breaks the 3-fold rotational symmetry (see Supplementary Information Note 4). Thus, this zone-folded ~30cm$^{-1}$ mode is a direct consequence of BTS (labeled as P$_{BTS}$), and its strength should scale with the primary zigzag AFM order parameter $\eta$.

Having established signatures of both BRS and BTS, as well as their relationship to the primary and composite order parameters, we proceed to examine their evolutions upon the dimensionality reduction. Figure 3d shows Raman spectra of P$_{BRS}$ and P$_{BTS}$ at 10K for a range of thickness from bulk down to 2L. The signature of BTS, P$_{BTS}$, vanishes when the thickness reduces below ~10.6nm, whereas the footprint for BRS, P$_{BRS}$, remains observable from bulk down to 2L NiPS$_3$. This contrast reveals the fact that the zigzag AFM order in 3D bulk NiPS$_3$ transitions into the nematic order in 2D few-layer NiPS$_3$ across a critical thickness of $t_C$ ~10.6nm. We further quantify the thickness dependence of both signatures, via four important parameters, the frequency separation of P$_{BRS}$ ($\Delta\omega_{BRS}$), the line width of P$_{BRS}$ ($\Gamma_{BRS}$), the linewidth of P$_{BTS}$ ($\Gamma_{BTS}$), and the intensity ratio between P$_{BTS}$ and P$_{BRS}$ ($I_{BTS}/I_{BRS}$). It has been discussed above that $\Delta\omega_{BRS}$ scales with the nematic order parameter, and its thickness independence in Fig. 3e confirms the robust presence of nematicity down to the bilayer. Furthermore, the nematicity coherence determines $\Gamma_{BRS}$, and the zigzag AFM coherence sets $\Gamma_{BTS}$. Figure 3f shows that $\Gamma_{BRS}$ remains nearly constant across $t_C$ ~10.6nm and only starts to increase slightly in 3L and more significantly in 2L. In contrast, Figure 3g illustrates a clearly divergent behavior of $\Gamma_{BTS}$ approaching $t_C$ from above. The distinct trends between $\Gamma_{BRS}$ and $\Gamma_{BTS}$ across $t_C$ clearly demonstrate the separation between the primary order (i.e., the zigzag AFM order of region I) and the vestigial order (i.e., the nematic order of region II) in the phase diagram illustrated in Fig. 1a. Finally, the suppression of $I_{BTS}/I_{BRS}$ down to zero when thinning NiPS$_3$ across $t_C$ further corroborates with the melting of the zigzag AFM order but the survival of the vestigial nematicity below $t_C$.

Having confirmed the presence of intrinsic vestigial nematicity in few-layer (thickness less than $t_C$) NiPS$_3$ belong to region II of Fig. 1a, we move forward to check the symmetry class of this nematicity. We select a large size 4L NiPS$_3$ (lateral dimension of ~40$\mu$m) to image the nematic domains using the scanning optical linear dichroism (LD) microscopy. The LD technique is known to be sensitive to BRS and has been used to study few-layer NiPS$_3$[41]. For the same reason as P$_{BRS}$ above, the LD signal scales with the vestigial nematic order parameter, rather than the zigzag AFM order parameter. Figure 4a shows a LD map for the 4L NiPS$_3$, where domains of different LD signals can be seen with clear boundaries between different domains. Surveying the angular dependence of the LD-induced polarization rotation across the 4L sample, we find that there are three, and only three, distinct patterns that are shown in Fig. 4b. These three patterns are related by the 3-fold rotational operation, corresponding to the three nematic domain states with the nematic order parameter rotated by 120º from one another i.e., $Z_3$ Potts-nematicity (schematic shown in the inset of Fig.4e-g). The *a* axis direction, defined as the nematic order parameter orientation, is calibrated by the polarization resolved photoluminescence (PL) measurement (see Supplementary Information Note 5). Such three Potts-nematic domain states are further confirmed by the angular dependence of the lower frequency $P_{BRS}$ mode (Figs. 4e-4f) whose anisotropy originates from the rotational symmetry reduction by the vestigial nematicity in the few-layer case. Thermal cycling does not change the nematic domains and nematic order parameter orientations within domains (see Supplementary Information Note 6), indicating that the nematic domain states are likely pinned by the structural monoclinic interlayer stacking. We note that the weak coupling between the structural monoclinicity and the vestigial nematicity should not impact the physics of interest here that enhanced fluctuations in 2D melt the zigzag AFM order and introduce the $Z_3$ Potts-nematicity.

To comprehend the experimental finding of the crossover from 3D zigzag AFM to 2D Potts-nematicity when the NiPS$_3$ thickness decreases across $t_c$, we further performed large-scale Monte Carlo simulations to visualize the magnetic phase for bilayer NiPS$_3$ and examine its primary ($\eta$) and composite ($\eta^\dagger \tau \eta$) order parameters. The magnetic state at finite temperatures is simulated under the spin Hamiltonian with intralayer Heisenberg exchange coupling up to the third nearest neighbor ($J_1, J_2, J_3$), a strong easy-plane anisotropy ($D_z$), a weak easy-axis anisotropy ($D_x$), and an isotropic interlayer exchange coupling ($J_{\text{inter}}$) (see Methods). Figure 5a plots the temperature dependence of the nematicity order parameter $\eta^\dagger \tau \eta = \langle |m_3| \rangle$ scaled with $L^{\beta/\nu}$ for four system sizes $L = 36, 60, 84, 108$, where temperature is in unit of $|J_1|$ and $\beta$ and $\nu$ are critical exponents for 2D $Z_3$ Potts-nematicity. All four traces cross at a single temperature (inset of Fig. 5a), defining the nematicity critical temperature, $T_{\text{N,2L}} = 1.285|J_1| = 67.1$ K, on the same order of the experimental value of 120K in Fig. 2d. This observation clearly confirms the emergence of Potts-nematicity in 2L NiPS$_3$. We further contrast this 2D Potts-nematic state against the zigzag AFM order by computing and comparing the nematicity and spin correlation, $C_{\text{nematicity}}(r)$ and $C_{\text{spin}}(r)$, at a temperature $T=1.25|J_1|<T_{\text{N,2L}}$, as shown in Fig. 5b. After the initial decay in short distance for both curves, which are normalized to their corresponding values at r = 2, $C_{\text{nematicity}}(r)$ reaches a saturation plateau that is consistent across all four system sizes, whereas in contrast, $C_{\text{spin}}(r)$ keeps having further decay as the system size increases. Note that the increase at distance greater than half of the system size is due to the finite size effect. The size-independent plateau behavior for $C_{\text{nematicity}}(r)$ confirms the long-range order of the composite nematic order parameter, but the size-dependent ever-decaying trend for $C_{\text{spin}}(r)$ demonstrate the "disorder" nature of the zigzag AFM state, corroborating with our experimental finding and belonging to Region II in Fig. 1a. To directly visualize the real-space spin arrangements for this unique magnetic phase, we plot a snapshot of the simulated result in Fig. 5c for spin texture and in Fig. 5d for normalized nematicity director texture. The spins (red and blue arrows) in Fig. 5c are randomly canted away from the zigzag chain direction, losing the spin coherence at long distance as revealed by Fig. 5b. As a comparison, the nematicity directors (green double arrows) in Fig. 5d show a much better homogeneity and therefore a long-range coherence as found in Fig. 5b.

Our work on the thickness dependence of spin fluctuations and magnetic phases in NiPS$_3$ clearly demonstrate that the enhanced fluctuations in 2D introduce the novel $Z_3$ Potts-nematic state by partially melting the conventional zigzag AFM order. Based on this result, we envision multiple emergent research opportunities as following. First, it is a promising venue to exploit enhanced fluctuations in the 2D, rather than suppressing them, to discover spin-fluctuation-driven novel magnetic phases, including and going beyond this $Z_3$ spin-induced Potts-nematicity. This direction distinguishes from the mainstream effort in 2D magnetism research thus far. Second, it remains as an open experimental question what the spin coherence length is in 2D NiPS$_3$, while the corresponding Potts-nematicity coherence length is much beyond optical wavelength (i.e., ~1$\mu$m). In fact, it is a more general question experimentally how partially melting the primary order parameter is in a vestigial state, to address which experimental techniques with high spatial resolution (i.e., ~ nanoscale) are required. Third, because the monoclinic interlayer stacking provides an effective field to couple with the Potts-nematicity in 2D NiPS$_3$, it would be interesting to explore what the magnetic state will be if twisting two NiPS$_3$ flakes to create a moiré superlattice whose supercell contains all three monoclinic stacking geometries. Finally, given that 3D NiPS$_3$ realizes the Mott insulating state with the strong electron correlations[64] and also hosts ultra-narrow exciton emission lines possibly from the many-body effect[40,65], 2D NiPS$_3$ offers a unique platform to explore the interplay between spin degree of freedom (DOF) with strong fluctuations and charge DOF of strong correlations, a new parameter regime inaccessible previously in 3D systems.

**Acknowledgements**

We acknowledge the valuable discussions with Rafael Fernandes and Jörn Venderbos. L.Z. acknowledges support by the NSF grant no. DMR-2103731, ONR grant no. N00014-21-1-2770 and the Gordon and Betty Moore Foundation Award GBMF10694. R.H. acknowledges support by the NSF grant no. DMR-2104036. C.D. acknowledges the support from U.S. Department of Energy (DOE), Office of Science, Basic Energy Sciences (BES), under award No. DE-SC0022946. Z.Y.M. acknowledges the Research Grants Council (RGC) of Hong Kong Special Administrative Region (SAR) of China (Projects Nos. 17301420, 17301721,

AoE/P-701/20, 17309822 and HKU C7037-22G), the ANR/RGC Joint Research Scheme sponsored by the RGC of Hong Kong SAR of China and French National Research Agency (Project No. A_HKU703/22). D.M. acknowledges support from the NSF grant no. DMR- 1808964. K.S. acknowledges support by ONR grant no. N00014-21-1-2770 and the Gordon and Betty Moore Foundation Award GBMF10694. Q.L. and H.D. acknowledge support by ONR grant no. N00014-21-1-2770 and the Gordon and Betty Moore Foundation Award GBMF10694.

**Author Contributions Statement**

Z.S. and L.Z. conceived the idea and initiated this project. Z.S. exfoliated NiPS$_3$ thin flakes with different layer numbers. G.Y., Z.Y., C. N. and Z.S. carried out the Raman experiments under the supervision of L.Z. and R.H.. M.H. performed the NV spin relaxometry under the supervision of C.R.D. C.Z. carried out the Monte Carlo simulations under the supervision of K.S. and Z.Y.M. Q. L. and Z. S. carried out the AFM measurement and the photoluminescence measurement with assistance of H.D.. R.X. and D.M. grew the high-quality NiPS$_3$ bulk single crystals. K.S. and Z.S. performed the theoretical analysis. Z.S., L.Z. and R.H. analyzed data and wrote the manuscript. All the authors participated in the discussion of the results.

**Competing Interests Statement**

The authors declare no competing interest.

**Data Availability**

All data that support the plots within this paper are available from the corresponding authors upon reasonable request.

**Figure Caption**

**Figure 1| Schematic phase diagrams of vestigial order and magnetic states in NiPS$_3$. a,** Schematic phase diagram to show the primary order (η) and the vestigial order ($\eta^\dagger \tau \eta$) as functions of temperature (vertical axis) and other independent factors, such as enhanced fluctuations at reduced dimensionality or disorder effect (horizontal axis). Three distinct regions are shown: (I) long-range primary order ($\langle \eta \rangle \neq 0$ and $\langle \eta^\dagger \tau \eta \rangle \neq 0$), (II) long-range vestigial order with short-range primary order ($\langle \eta \rangle = 0$ but $\langle \eta^\dagger \tau \eta \rangle \neq 0$),

and (III) disorder ($\langle\eta\rangle = 0$ and $\langle\eta^\dagger\tau\eta\rangle = 0$). Examples of primary order *v.s.* vestigial order are listed in the phase diagram, including charge or spin density wave v.s. nematicity[4,21,22], pair density wave v.s. charge-$4e$ superconductivity or nematicity or charge density wave[23-26], superconductivity v.s. charge-$4e$ superconductivity or nematicity[27-29], *etc*. **b,** Schematic of the long-range zigzag AFM order in 3D NiPS$_3$ with both BRS and BTS, corresponding to the region I in **a**. The Ni$^{2+}$ spins are aligned along the zigzag chain. The green parallelogram represents the in-plane unit cell of the crystal structure without the AFM order, and the blue rectangle shows the doubled in-plane unit cell for the zigzag AFM state. Inset: Brillouin zone (BZ) folding in the momentum space. The *M* point at BZ boundary without the AFM order folds to the $\Gamma$ point of BZ center by the zigzag AFM order. **c,** Schematic of the spin-induced nematic state in 2D NiPS$_3$ with BRS but without BTS, corresponding to the region II in **a**. Inset: Corresponding BZ without folding due to the lack of BTS. **d,** Schematic of the spin disordered state without BRS and BTS in NiPS$_3$, corresponding to the region III in **a**. Inset: Corresponding BZ. In **b-d**, only the Ni atoms are shown for simplicity.

**Figure 2| Thickness-dependent spin fluctuations in few-layer NiPS$_3$ measured by NV relaxometry and Raman spectroscopy. a,** Temperature dependence of NV spin relaxation rate $\Gamma_\text{M}$ measured at NV centers underneath the 10 nm- and 4L-NiPS$_3$. Inset: NV spin relaxation map for the few-layer NiPS$_3$ samples at 170 K. The black dashed lines outline the boundary of the 10 nm- and 4L-NiPS$_3$ samples. The scale bar is 4 μm. **b,** Fitted magnetic susceptibility $\chi$ of 10 nm- and 4L-NiPS$_3$ as a function of temperature. **c,** Normalized QES Raman spectra measured with λ=532 nm in 13.1nm and 4L-NiPS$_3$ at 170 K, respectively. The shaded area denotes the breathing mode in 4L-NiPS$_3$ that is excluded in calculating the QES spectral integral in **d**. **d,** Temperature-dependent integrated QES intensity over 8 – 40 cm$^{-1}$ in NiPS$_3$ with different thickness.

**Figure 3| Thickness dependence of BTS and BRS signatures in Raman spectroscopy. a,** Linearly polarized Raman spectra measured with λ=633 nm in both parallel (black curves) and crossed (gray curves) channels at 10 K. The blue and red areas denote the P$_\text{BTS}$ mode and P$_\text{BRS}$ mode, respectively. **b,** Temperature-dependent Raman spectra of P$_\text{BTS}$ mode in the parallel channel. Here, to have a clearer presentation of the temperature dependence of P$_\text{BTS}$ mode, we subtracted the QES signal from the raw data above 90 K. **c,** Temperature-dependent Raman spectra of P$_\text{BRS}$ mode in the parallel channel. **d,** Layer-dependent Raman spectra in both parallel (filled circles) and crossed (empty circles) channels at *T*=10 K. The dark (light) solid line represents the fitted curves in the parallel (crossed) channel by using the Lorentz function. **e,** Thickness dependence of fitted frequency separation of P$_\text{BRS}$, $\Delta\omega_\text{BRS} = \omega_\text{BRS-h} - \omega_\text{BRS-l}$, at *T*=10 K. Here, $\omega_\text{BRS-h}$ was fitted from cross channel and $\omega_\text{BRS-l}$ was fitted from parallel channel. **f,** Thickness dependence of fitted linewidth $\Gamma_\text{BRS}$ of P$_\text{BRS}$ at *T*=10 K. Here, $\Gamma_\text{BRS}$ was fitted from the low-frequency mode of split P$_\text{BRS}$ mode in parallel channel. **g,** Thickness dependence of fitted linewidth $\Gamma_\text{BTS}$ of P$_\text{BTS}$ at *T*=10 K. The linewidth shows a divergent behavior as the thickness decreasing to the critical thickness ~ 10 nm **h,**

Thickness dependence of the fitted intensity ratio of $I_{BTS}/I_{BRS}$ at $T=10$ K. Here we choose the intensity ratio of $I_{BTS}/I_{BRS}$, which is defined as $(I_{BTS\text{-parallel}}+I_{BTS\text{-crossed}})/(I_{BRS\text{-l-parallel}}+I_{BRS\text{-l-crossed}})$, to focus on the relative strength between $P_{BTS}$ and $P_{BRS}$ and exclude the possible impacts of the absolute intensity of $P_{BTS}$ and $P_{BRS}$ from different dielectric environment. The color shading in **e-h** corresponding to region II (yellow) and I (blue), following that in Fig. 1a.

**Figure 4|Polarization-dependent LD and Raman data in 4L-NiPS$_3$. a,** Spatially resolved LD map in a 4L-NiPS$_3$. The black dashed lines outline the boundary of 4L-NiPS$_3$ samples (optical image shown in Supplementary Fig. S6j). The three dots denote the three Potts-nematic domains with the nematic order parameter rotated 120° from one another. **b-d,** Angular-dependent polarization rotation $\Delta\theta$ in the corresponding nematic domains in the 4L-NiPS$_3$ as shown in **a**, in which red area is positive sign and gray area is negative sign. The arrows indicate the $a$ axis in the corresponding domain, which is determined by the polarization-dependent PL measurement (Supplementary Note 5). **e-g,** Polar plots of the polarization dependence of fitted $P_{BRS\text{-l}}$ intensity at $T=10$ K in the corresponding Potts-nematic domain in the 4L-NiPS$_3$ shown in **a**, in which the arrows denote the $a$ axis in the corresponding domain. Inset: Schematic of the nematic order directors for the three different Potts-nematic domain states.

**Figure 5|Monte Carlo calculations of the magnetic state in 2L NiPS$_3$. a,** Temperature dependence of the scaled Potts-nematic order parameter ($\langle|m_3|\rangle$) for four different system sizes, L = 36, 60, 84, and 108. Inset: Zoom-in around nematic phase transition temperature $T_{N,\,2L}=1.285|J_1|$. The crossing point of scaled nematic order parameter for different system size determines the nematic transition point with the 2D $Z_3$ Potts exponents $\beta = 1/9$, $\nu = 5/6$. **b,** Correlation of Potts-nematicity ($C_{nematicity}$) and spin ($C_{spin}$) at a temperature $T=1.25|J_1|<T_{N,2L}$, as a function of distance for four different system sizes, respectively. The plot is in log-log scale and we normalize the two correlation functions at r = 2. **c,** Snapshot of simulated spin configuration below $T_{N,\,2L}$, which shows the lack of spin coherence. **d,** Image of simulated nematic director (normalized $e_3$ configuration) for the spin state in **c**, showing homogeneity of nematicity.

**Methods**

**Sample fabrication.** High-quality NiPS$_3$ single crystals were grown by the chemical vapor transport (CVT) method[66]. Few-layer NiPS$_3$ samples were prepared on Si/SiO$_2$ by mechanical exfoliation from high-quality bulk NiPS$_3$ single crystals in the nitrogen-filled glovebox with oxygen level below 0.1 ppm and water level below 0.5 ppm. For the measurement of NV magnetometry, one hBN thin flake was first picked up by the poly(bisphenol A carbonate) stamp, and then the selected few-layer NiPS$_3$ samples on Si/SiO$_2$ were picked up by the hBN flake on the poly(bisphenol A carbonate) stamp. The hBN/few-layer NiPS$_3$ was transferred onto diamond membrane for NV magnetometry measurements. The thickness of the few-layer samples were first determined based on the optical contrast and then confirmed by the atomic force microscopy (AFM) after measurement.

**Raman measurement.** Raman spectroscopy measurements were performed by using the λ=532 nm and 633 nm excitation lasers. The laser beam on the sample was focused down to ~2–3 μm in diameter, and the laser power was kept below 100 μW to minimize the local heating effect during measurements. Backscattering geometry was used. The scattered light was detected by a thermoelectrically cooled charge-coupled device (CCD) camera from Horiba Scientific, with a spectral resolution of 0.4 cm$^{-1}$. The temperature dependent Raman measurement was carried out in a commercial variable-temperature, closed-cycle cryostat (Cryo Industries of America). For the magnetic field dependent Raman measurement, a commercial superconducting magnet (Cryo Industries of America) was used to achieve magnetic field from 0 to 7 T.

**Raman spectra fitting procedure.** We used multiple Lorentzian functions, $\sum_N \frac{A_N (\frac{\Gamma_N}{2})^2}{(\omega-\omega_N)^2+(\frac{\Gamma_N}{2})^2} + C$, to fit the P$_{BTS}$ and P$_{BRS}$ modes, where $A_N$ is the peak intensity, $\omega_N$ is the peak position, $\Gamma_N$ is the peak linewidth, $C$ is background, $N=1$ for the single peak and $N=2$ for the double peaks. For the QES signal, we first remove the breathing mode signal if present and then fitted the Raman data in the Raman shift range 8-40 cm$^{-1}$ by using the Lorentzian function centered at 0 cm$^{-1}$, $\frac{A_N (\frac{\Gamma_N}{2})^2}{\omega^2+(\frac{\Gamma_N}{2})^2} + C$ where $C$ is a constant background. Finally, we integrated the fitted curves without background in range 8-40 cm$^{-1}$ as the QES intensity.

**Derivation of angular dependence of Raman modes.** We derived the polarization dependence of Raman intensity in the parallel channel as a function of polarization rotation angle $\theta$. The measured Raman intensity ($I$) is proportional to the square of the product of incident light polarization (ê$_i$), Raman tensor (R) and the scattered light polarization (ê$_f$): $I \propto |\langle \hat{e}_i | R | \hat{e}_f \rangle|^2$. In the parallel channel, the incident light $|\hat{e}_i\rangle = \begin{pmatrix} cos\theta \\ sin\theta \end{pmatrix}$, and the scattered light $|\hat{e}_f\rangle = \begin{pmatrix} cos\theta \\ sin\theta \end{pmatrix}$. The Raman tensor of P$_{BTS}$ mode is $\begin{pmatrix} a & 0 \\ 0 & b \end{pmatrix}$. So, the intensity $I_{BTS} \propto |(cos\theta, sin\theta) \begin{pmatrix} a & 0 \\ 0 & b \end{pmatrix} \begin{pmatrix} cos\theta \\ sin\theta \end{pmatrix}|^2 = \frac{1}{8}(a-b)^2 cos4\theta + \frac{1}{4}(a^2-b^2)cos2\theta + \frac{1}{8}(3a^2+2ab+3b^2)$ (Supplementary Fig. S4a). The Raman tensors of P$_{BRS-l}$ and P$_{BRS-h}$ are $\begin{pmatrix} e & 0 \\ 0 & -e \end{pmatrix}$ and $\begin{pmatrix} 0 & f \\ f & 0 \end{pmatrix}$, respectively. So the intensity $I_{BRS-l} \propto |(cos\theta, sin\theta) \begin{pmatrix} e & 0 \\ 0 & -e \end{pmatrix} \begin{pmatrix} cos\theta \\ sin\theta \end{pmatrix}|^2 = \frac{e^2}{2}(\cos 4\theta + 1)$ and $I_{BRS-l} \propto$

$|(cos\theta, sin\theta)\begin{pmatrix} 0 & f \\ f & 0 \end{pmatrix}\begin{pmatrix} cos\theta \\ sin\theta \end{pmatrix}|^2 = \frac{f^2}{2}(1 - \cos 4\theta) = \frac{f^2}{2}\left(\cos 4\left(\theta + \frac{\pi}{8}\right) + 1\right)$. Both patterns of $P_{BRS-l}$ and $P_{BRS-h}$ show four-fold symmetry and have a 45° phase shift from one another (Supplementary Fig. S4b).

**NV spin relaxometry measurement.** The widefield NV spin relaxation rate measurements were performed in a closed-cycle optical cryostat, and the measurement temperature ranges from 4.5 K to 350 K. A microsecond long green laser pulse generated by an electrical driven 515 nm laser is used to initialize NV centers to $m_s = 0$ state and read out the spin-state dependent photoluminescence. The laser beam spot focused on the diamond surface is around 20 μm ×20 μm. A CMOS camera is used to collect NV fluorescence images, and both pulses to trigger the camera exposures and drive the laser are controlled by a programmable pulse generator. Nanosecond-long microwave pulses were generated by sending the continuous microwave currents to a microwave switch electrically controlled by a programmable pulse generator, and then were delivered to Au stripline patterned on diamond samples. An external field ~ 70 G is generated by a cylindrical NdFeB permanent magnet attached to a scanning stage inside the optical cryostat. Further details of NV relaxometry measurement protocol can be found in Supplementary Note 1.

**Linear dichroism measurement.** LD measurement was performed by using laser diode with wavelength 635 nm (Thorlabs PL202) and 20 μW power. A photo-elastic modulator (PEM; PEM-100, Hinds Instruments) on the incident path was used for the measurement of LD. The incident light with linear polarization with 45° angle with respect to the PEM fast axis was modulated by a mechanical copper and PEM with retardance of λ/2 by sequence. The light was then passed through a half-wave plate and focused onto the sample in normal incidence with a 50× objective lens. The reflected light passed the half-wave plate again and then went into the detector through a plate beamsplitter. The reflected light was measured by a balanced amplified photodetector (ThorLabs PDB210A). The signal was demodulated by two lock-in amplifiers (Zurich Instruments, MFLI 500kHz/5MHz) separately, of which one lock-in amplifier is referenced to the second harmonic of the PEM frequency 2f = 84.183 kHz and the other lock-in amplifier is referenced to the chopping frequency f = 511 Hz to get the total reflectance as a normalization. For the polarization-dependent LD measurement, the polarization direction of incident light was rotated by the half-wave plate. Thus, the polarization rotation $\Delta\theta$ induced by the LD of few-layer $NiPS_3$ can be written as $\Delta\theta = LDsin(2\alpha + \varphi)$, in which $\alpha$ is the polarization direction of incident light, and $\varphi$ is phase dependent on the relative angle between the crystal axis and the polarization direction of incident light at zero. $\Delta\theta$ is zero when the polarization direction of incident light is along the crystal axes direction. The angular dependent signal was fitted by the equation $\Delta\theta = LDsin(2\alpha + \varphi)$ and a small nonzero background from inevitable birefringence of the optical components in the whole setup was subtracted during the fitting process. For the LD scanning measurement, a two-axis Galvo scanning mirror paired with a confocal imaging system was used to move the light spot on the sample in normal incidence with a constant fluence and the half-wave plate was fixed at a certain angle during scanning.

**Polarization-dependent PL measurement.** PL measurement was performed by using a continuous-wave solid-state laser with 532 nm (Thorlabs) to excite the few-layer NiPS$_3$ sample. The beam size focused on sample was ~2.5 μm in diameter. The collected signals are detected by a Princeton Instruments spectrometer with a CCD camera. The sample is kept at 5 K using a Montana CA100 system. For the polarization-dependent PL measurements, a polarizer was put in the detection beam line to change the polarization direction of collected light into the CCD camera.

**Monte Carlo simulations.** The magnetic Hamiltonian of NiPS$_3$ can be described by 2D XY type spin Hamiltonian with an easy-plane single-ion anisotropy ($D_z > 0$) and an easy-axis single-ion anisotropy ($D_x < 0$)[48]:

$$H = \frac{1}{2} \sum_{\langle ij \rangle} J_{ij}(S_i^x S_i^x + S_i^y S_j^y + S_i^z S_j^z) + D_z \sum_i (S_i^z)^2 + D_x \sum_i (S_i^x)^2$$

In which $J_{ij}$ is exchange coupling parameters, $D_z$ is easy-plane single-ion anisotropy and $D_x$ is easy-axis single-ion anisotropy. Here, we considered the nearest exchange coupling interaction $J_1$, next-nearest exchange coupling interaction $J_2$, third-nearest exchange coupling interaction $J_3$ and exchange coupling interaction between neighboring layers $J_{\text{inter}}$. We performed large-scale Monte Carlo simulations on bilayer NiPS$_3$ systems. In our simulations, we employed various techniques including local updates, Woff updates[67], and the over-relaxation technique[68]. The spins are treated as classical O(3) vectors, and the simulation is warmed up with $10^4$ Monte Carlo steps. In our simulation, we set the parameters $\{J_1, J_2, J_3, J_{\text{inter}}, D_x, D_z\}$ to be $\{-1, 0, 2.22, -0.07, -0.002, 0.047\}$, with $|J_1|$ as the unit[48,69]. We simulated the bilayer case ($L_z = 2$) with periodic boundary conditions along the $x$ and $y$ directions, but with open boundary conditions along the $z$ direction, where the linear system size ranges from $L = 36$ to $L = 108$.

In our simulation, we define the order parameter of the $Z_3$ rotational symmetry $m_3$ as

$$m_3 = \sigma_1 \epsilon_1 + \sigma_2 \epsilon_2 + \sigma_3 \epsilon_3$$

Here, $\sigma_i$ represents the bond direction, given by $\sigma_1 = (0,1)$, $\sigma_2 = (-\sqrt{3}/2, -1/2)$, and $\sigma_3 = (\sqrt{3}/2, -1/2)$. And $\epsilon_i$ refers to the average neighboring bond energy along the $\sigma_i$ direction, i.e., $\epsilon_i = \mathbf{S}_r \cdot \mathbf{S}_{r+e_i}$. This definition of the nematic order parameter is consistent with $\eta^\dagger \tau \eta = \left(\frac{\sqrt{3}}{2}(\eta_3^2 - \eta_2^2), \frac{1}{2}(2\eta_1^2 - \eta_2^2 - \eta_3^2)\right)$ via writing $\eta^\dagger \tau \eta$ as $\eta_1^2 \boldsymbol{\sigma_1} + \eta_2^2 \boldsymbol{\sigma_2} + \eta_3^2 \boldsymbol{\sigma_3}$ where $\boldsymbol{\sigma_j} = \left(\cos[\frac{2\pi}{3}(j-1) + \frac{\pi}{2}], \sin[\frac{2\pi}{3}(j-1) + \frac{\pi}{2}]\right)$. Noting that $\eta_j^2 = \eta_j \cdot \eta_j = (S_r - S_{r+e_j}) \cdot (S_r - S_{r+e_j}) = -2 S_r \cdot S_{r+e_j} + S_r^2 + S_{r+e_j}^2$ where the last two terms $S_r^2$ and $S_{r+e_j}^2$ are constants, $\eta_1^2 \sigma_1 + \eta_2^2 \sigma_2 + \eta_3^2 \sigma_3 = -2 \sum_j (S_r \cdot S_{r+e_j}) \sigma_j + const \sum_j \sigma_j = -2 \sum_j (S_r \cdot S_{r+e_j}) \sigma_j$ because of $\sum_j \sigma_j = 0$. Therefore, $\eta^\dagger \tau \eta = \left( \frac{\sqrt{3}}{2}(\eta_3^2 - \eta_2^2), \frac{1}{2}(2\eta_1^2 - \eta_2^2 - \eta_3^2) \right)$ and $m_3$ only differ from each other by a constant factor of 2.

The norm of $m_3$, $\langle |m_3| \rangle$, is the nematic order parameter. If $Z_3$ rotation symmetry is present, $\langle |m_3| \rangle$ will vanish, however, $\langle |m_3| \rangle$ will become non-zero in the case of $Z_3$ rotation symmetry breaking. Meanwhile, we measure the nematic correlation, $C_{e_3}(r) = \langle e_{3,i} e_{3,i+r} \rangle$, in which the nematic director $e_3 = \sigma_1 \epsilon_{i,1} + \sigma_2 \epsilon_{i,2} + \sigma_3 \epsilon_{i,3}$ defined on each lattice site. By examining the behavior of $\langle |m_3| \rangle$ and $C_{e_3}(r)$, we can identify the occurrence of nematic order. We also measure the spin correlation, $C_s(r) = \langle S_i S_{i+r} \rangle$, to investigate whether long-range spin order occurs.

**References**

1. Chaikin, P. M., Lubensky, T. C. & Witten, T. A. *Principles of Condensed Matter Physics*. Vol. 10 (Cambridge university press Cambridge, 1995).
2. Cardy, J. *Scaling and renormalization in statistical physics*. Vol. 5 (Cambridge university press, 1996).
3. Nie, L., Tarjus, G. & Kivelson, S. A. Quenched disorder and vestigial nematicity in the pseudogap regime of the cuprates. *Proceedings of the National Academy of Sciences* **111**, 7980-7985 (2014).
4. Fernandes, R. M., Orth, P. P. & Schmalian, J. Intertwined vestigial order in quantum materials: Nematicity and beyond. *Annual Review of Condensed Matter Physics* **10**, 133-154 (2019).
5. Huang, B. *et al.* Layer-dependent ferromagnetism in a van der Waals crystal down to the monolayer limit. *Nature* **546**, 270-273 (2017).
6. Gong, C. *et al.* Discovery of intrinsic ferromagnetism in two-dimensional van der Waals crystals. *Nature* **546**, 265-269 (2017).
7. Lee, K. *et al.* Magnetic order and symmetry in the 2D semiconductor CrSBr. *Nano Letters* **21**, 3511-3517 (2021).
8. Lee, J.-U. *et al.* Ising-type magnetic ordering in atomically thin $FePS_3$. *Nano letters* **16**, 7433-7438 (2016).
9. Ni, Z. *et al.* Imaging the Néel vector switching in the monolayer antiferromagnet $MnPSe_3$ with strain-controlled Ising order. *Nature Nanotechnology* **16**, 782-787 (2021).
10. Kivelson, S. A., Fradkin, E. & Emery, V. J. Electronic liquid-crystal phases of a doped Mott insulator. *Nature* **393**, 550-553 (1998).
11. Ando, Y., Segawa, K., Komiya, S. & Lavrov, A. Electrical resistivity anisotropy from self-organized one dimensionality in high-temperature superconductors. *Physical Review Letters* **88**, 137005 (2002).
12. Hinkov, V. *et al.* Electronic liquid crystal state in the high-temperature superconductor $YBa_2Cu_3O_6$. 45. *Science* **319**, 597-600 (2008).
13. Lawler, M. *et al.* Intra-unit-cell electronic nematicity of the high-$T_c$ copper-oxide pseudogap states. *Nature* **466**, 347-351 (2010).
14. Achkar, A. *et al.* Nematicity in stripe-ordered cuprates probed via resonant x-ray scattering. *Science* **351**, 576-578 (2016).
15. de La Cruz, C. *et al.* Magnetic order close to superconductivity in the iron-based layered $LaO_{1-x}F_xFeAs$ systems. *Nature* **453**, 899-902 (2008).

16  Chu, J.-H., Kuo, H.-H., Analytis, J. G. & Fisher, I. R. Divergent nematic susceptibility in an iron arsenide superconductor. *Science* **337**, 710-712 (2012).

17  Fang, C., Yao, H., Tsai, W.-F., Hu, J. & Kivelson, S. A. Theory of electron nematic order in LaFeAsO. *Physical Review B* **77**, 224509 (2008).

18  Xu, C., Müller, M. & Sachdev, S. Ising and spin orders in the iron-based superconductors. *Physical Review B* **78**, 020501 (2008).

19  Domröse, T. *et al.* Light-induced hexatic state in a layered quantum material. *Nature Materials*, 1-7 (2023).

20  Cho, C.-w. *et al.* $Z_3$-vestigial nematic order due to superconducting fluctuations in the doped topological insulators $Nb_xBi_2Se_3$ and $Cu_xBi_2Se_3$. *Nature communications* **11**, 3056 (2020).

21  Fradkin, E., Kivelson, S. A. & Tranquada, J. M. Colloquium: Theory of intertwined orders in high temperature superconductors. *Reviews of Modern Physics* **87**, 457 (2015).

22  Fernandes, R., Chubukov, A. & Schmalian, J. What drives nematic order in iron-based superconductors? *Nature Physics* **10**, 97-104 (2014).

23  Berg, E., Fradkin, E. & Kivelson, S. A. Charge-4e superconductivity from pair-density-wave order in certain high-temperature superconductors. *Nature Physics* **5**, 830-833 (2009).

24  Agterberg, D. F. *et al.* The physics of pair-density waves: Cuprate superconductors and beyond. *Annual Review of Condensed Matter Physics* **11**, 231-270 (2020).

25  Lee, P. A. Amperean pairing and the pseudogap phase of cuprate superconductors. *Physical Review X* **4**, 031017 (2014).

26  Berg, E., Fradkin, E., Kivelson, S. A. & Tranquada, J. M. Striped superconductors: how spin, charge and superconducting orders intertwine in the cuprates. *New Journal of Physics* **11**, 115004 (2009).

27  Jian, S.-K., Huang, Y. & Yao, H. Charge-4e Superconductivity from Nematic Superconductors in Two and Three Dimensions. *Physical Review Letters* **127**, 227001 (2021).

28  Fernandes, R. M. & Fu, L. Charge-4e superconductivity from multicomponent nematic pairing: Application to twisted bilayer graphene. *Physical Review Letters* **127**, 047001 (2021).

29  Hecker, M., Willa, R., Schmalian, J. & Fernandes, R. M. Cascade of vestigial orders in two-component superconductors: Nematic, ferromagnetic, s-wave charge-4e, and d-wave charge-4e states. *Physical Review B* **107**, 224503 (2023).

30  Wang, Y.-C., Yan, Z., Wang, C., Qi, Y. & Meng, Z. Y. Vestigial anyon condensation in kagome quantum spin liquids. *Physical Review B* **103**, 014408 (2021).

31  Daou, R. *et al.* Broken rotational symmetry in the pseudogap phase of a high-$T_c$ superconductor. *Nature* **463**, 519-522 (2010).

32    Sato, Y. *et al.* Thermodynamic evidence for a nematic phase transition at the onset of the pseudogap in YBa$_2$Cu$_3$O$_y$. *Nature Physics* **13**, 1074-1078 (2017).

33    Böhmer, A. E., Chu, J.-H., Lederer, S. & Yi, M. Nematicity and nematic fluctuations in iron-based superconductors. *Nature Physics* **18**, 1412-1419 (2022).

34    Kasahara, S. *et al.* Electronic nematicity above the structural and superconducting transition in BaFe$_2$(As$_{1-x}$P$_x$)$_2$. *Nature* **486**, 382-385 (2012).

35    Wu, F.-Y. The potts model. *Reviews of Modern Physics* **54**, 235 (1982).

36    Fernandes, R. M. & Venderbos, J. W. Nematicity with a twist: Rotational symmetry breaking in a moiré superlattice. *Science Advances* **6**, eaba8834 (2020).

37    Christensen, M. H., Birol, T., Andersen, B. M. & Fernandes, R. M. Theory of the charge density wave in AV$_3$Sb$_5$ kagome metals. *Physical Review B* **104**, 214513 (2021).

38    Little, A. *et al.* Three-state nematicity in the triangular lattice antiferromagnet Fe$_{1/3}$NbS$_2$. *Nature Materials* **19**, 1062-1067 (2020).

39    Ni, Z., Huang, N., Haglund, A. V., Mandrus, D. G. & Wu, L. Observation of giant surface second-harmonic generation coupled to nematic orders in the van der Waals antiferromagnet FePS$_3$. *Nano Letters* **22**, 3283-3288 (2022).

40    Kang, S. *et al.* Coherent many-body exciton in van der Waals antiferromagnet NiPS$_3$. *Nature* **583**, 785-789 (2020).

41    Hwangbo, K. *et al.* Highly anisotropic excitons and multiple phonon bound states in a van der Waals antiferromagnetic insulator. *Nature Nanotechnology* **16**, 655-660 (2021).

42    Wang, X. *et al.* Spin-induced linear polarization of photoluminescence in antiferromagnetic van der Waals crystals. *Nature Materials* **20**, 964-970 (2021).

43    Kim, K. *et al.* Suppression of magnetic ordering in XXZ-type antiferromagnetic monolayer NiPS$_3$. *Nature Communications* **10**, 345 (2019).

44    Dirnberger, F. *et al.* Spin-correlated exciton–polaritons in a van der Waals magnet. *Nature Nanotechnology* **17**, 1060-1064 (2022).

45    Wildes, A. R. *et al.* Magnetic structure of the quasi-two-dimensional antiferromagnet NiPS$_3$. *Physical Review B* **92**, 224408 (2015).

46    Lançon, D., Ewings, R., Guidi, T., Formisano, F. & Wildes, A. Magnetic exchange parameters and anisotropy of the quasi-two-dimensional antiferromagnet NiPS$_3$. *Physical Review B* **98**, 134414 (2018).

47    Joy, P. & Vasudevan, S. Magnetism in the layered transition-metal thiophosphates MPS$_3$ (M= Mn, Fe, and Ni). *Physical Review B* **46**, 5425 (1992).

48    Wildes, A. *et al.* Magnetic dynamics of NiPS$_3$. *Physical Review B* **106**, 174422 (2022).

49  Scheie, A. *et al.* Spin wave Hamiltonian and anomalous scattering in NiPS$_3$. *Physical Review B* **108**, 104402 (2023).

50  Doherty, M. W. *et al.* The nitrogen-vacancy colour centre in diamond. *Physics Reports* **528**, 1-45 (2013).

51  Rondin, L. *et al.* Magnetometry with nitrogen-vacancy defects in diamond. *Reports on Progress in Physics* **77**, 056503 (2014).

52  Van der Sar, T., Casola, F., Walsworth, R. & Yacoby, A. Nanometre-scale probing of spin waves using single electron spins. *Nature Communications* **6**, 7886 (2015).

53  Du, C. *et al.* Control and local measurement of the spin chemical potential in a magnetic insulator. *Science* **357**, 195-198 (2017).

54  Flebus, B. & Tserkovnyak, Y. Quantum-impurity relaxometry of magnetization dynamics. *Physical Review Letters* **121**, 187204 (2018).

55  Ku, M. J. *et al.* Imaging viscous flow of the Dirac fluid in graphene. *Nature* **583**, 537-541 (2020).

56  Kubo, R. The fluctuation-dissipation theorem. *Reports on Progress in Physics* **29**, 255 (1966).

57  Lemmens, P., Güntherodt, G. & Gros, C. Magnetic light scattering in low-dimensional quantum spin systems. *Physics Reports* **375**, 1-103 (2003).

58  Reiter, G. Light scattering from energy fluctuations in magnetic insulators. *Physical Review B* **13**, 169 (1976).

59  Halley, J. Light scattering as a probe of dynamical critical properties of antiferromagnets. *Physical Review Letters* **41**, 1605 (1978).

60  Brya, W. & Richards, P. M. Frequency moments for two-spin light scattering in antiferromagnets. *Physical Review B* **9**, 2244 (1974).

61  Richards, P. M. & Brya, W. Spin-fluctuation light scattering at high temperature. *Physical Review B* **9**, 3044 (1974).

62  Hashemi, A., Komsa, H.-P., Puska, M. & Krasheninnikov, A. V. Vibrational properties of metal phosphorus trichalcogenides from first-principles calculations. *The Journal of Physical Chemistry C* **121**, 27207-27217 (2017).

63  Kuo, C.-T. *et al.* Exfoliation and Raman spectroscopic fingerprint of few-layer NiPS$_3$ van der Waals crystals. *Scientific Reports* **6**, 20904 (2016).

64  Kim, S. Y. *et al.* Charge-spin correlation in van der Waals antiferromagnet NiPS$_3$. *Physical Review Letters* **120**, 136402 (2018).

65  Kim, D. S. *et al.* Anisotropic Excitons Reveal Local Spin Chain Directions in a van der Waals Antiferromagnet. *Advanced Materials* **35**, 2206585 (2023).

66      Matsuoka, T. *et al.* Pressure-Induced Insulator–Metal Transition in Two-Dimensional Mott Insulator NiPS3. *Journal of the Physical Society of Japan* **90**, 124706 (2021).

67      Wolff, U. Collective Monte Carlo updating for spin systems. *Physical Review Letters* **62**, 361 (1989).

68      Creutz, M. Overrelaxation and monte carlo simulation. *Physical Review D* **36**, 515 (1987).

69      DiScala, M. *et al.* Dimensionality dependent electronic structure of the exfoliated van der Waals antiferromagnet NiPS$_3$. *arXiv preprint arXiv:2302.07910* (2023).

# Figure 1

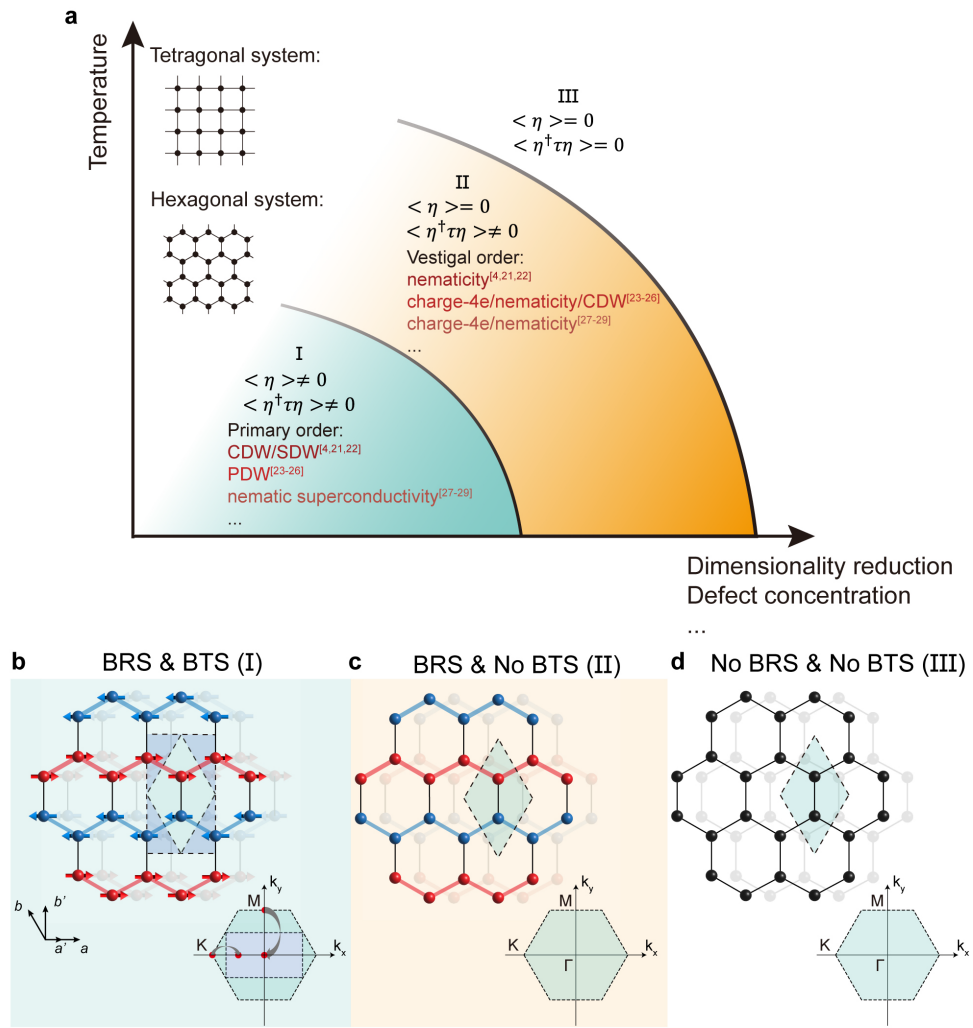

**Figure 2**

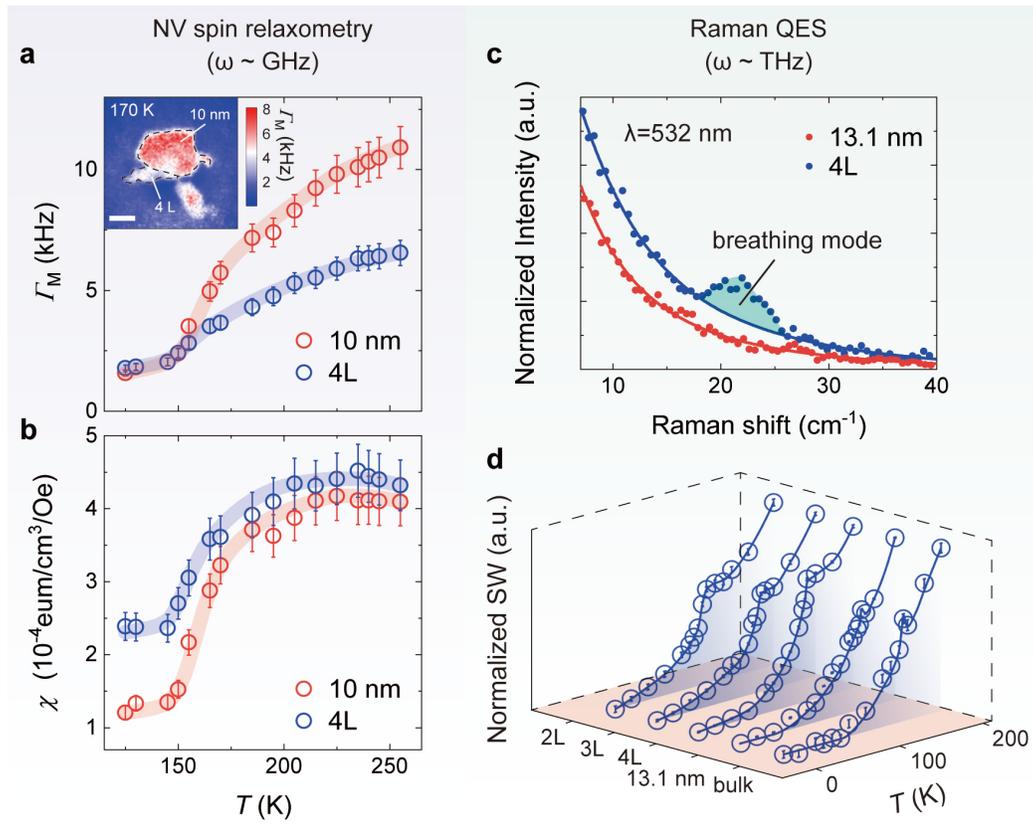

# Figure 3

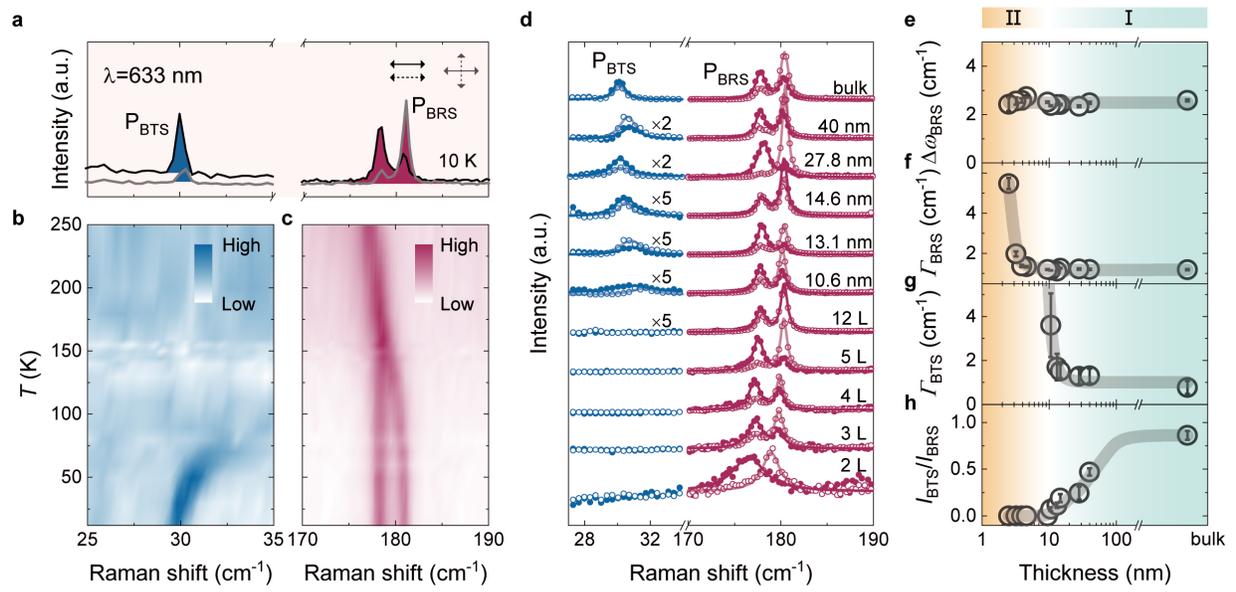

# Figure 4

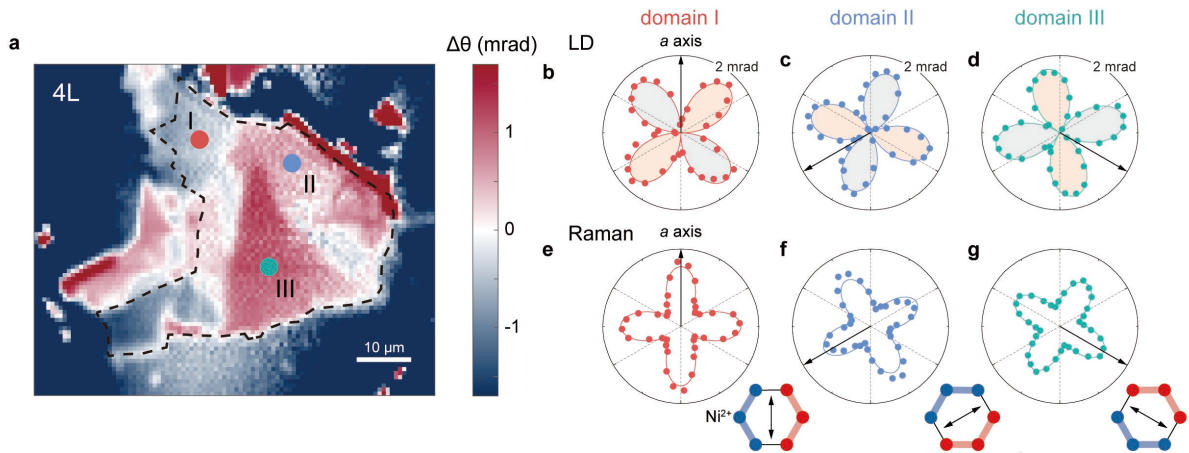

**Figure 5**

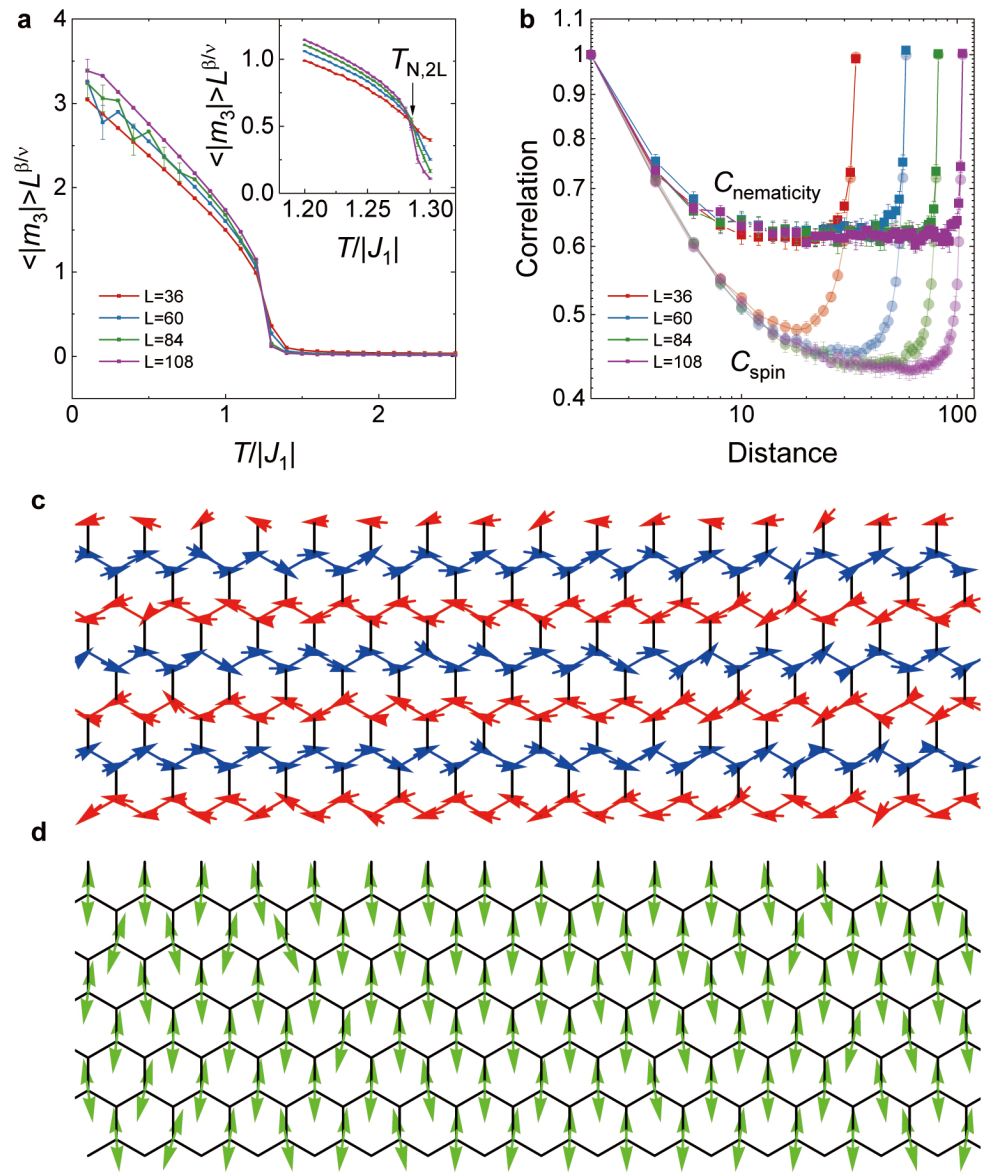